\documentclass[pra,final,onecolumn,showpacs,superscriptaddress,aps]{revtex4-1}
\usepackage{graphicx}
\usepackage{amssymb}
\usepackage{amsmath}
\usepackage{color}
\usepackage{multirow}
\usepackage{afterpage}
\usepackage{dsfont}
\begin{document}

\title{ Quantum Correlation and Coherence in a Mononuclear Nickel-Based Molecular Magnet}
\author{S. Bhuvaneswari}
\affiliation{Centre for Nonlinear Science (CeNSc), Department of Physics, Government College for Women (Autonomous), Kumbakonam  612001, Tamil Nadu, India}

\author{R. Muthuganesan}
\email{Corresponding author: rajendramuthu@gmail.com}
\affiliation{Department of Physics and Nanotechnology, SRM Institute of Science and
Technology, Kattankulathur -- 603203, Tamil Nadu, India}

\author{R. Radha}
\affiliation{Centre for Nonlinear Science (CeNSc), Department of Physics,  Government College for Women (Autonomous), Kumbakonam  612001, Tamil Nadu, India}

\begin{abstract}
We investigate the behaviors of thermal entanglement, quantum correlation beyond entanglement namely, measurement-induced nonlocality (MIN) and coherence in a nickel radical molecular magnet $(\text{Et}_3\text{NH})[\text{Ni(hfac)}_2L]$, whose spin-spin interactions are well described by the Heisenberg model. Using experimentally estimated coupling parameters, we compute the thermal state of the system and analyze the dependence of quantum resources on temperature and magnetic field.  The results indicate that the quantum resources of the nickel-radical molecular magnet persist even at room temperature. We show that while negativity (the entanglement measure) rapidly vanishes with increasing temperature and magnetic field, measurement-induced nonlocality and quantum coherence remain comparatively more stable and persist in regions where entanglement is absent. These results highlight the significance of nonclassical correlations beyond entanglement in thermally activated spin systems and suggest that such molecular magnets could serve as viable platforms for quantum information processing in realistic conditions.
\end{abstract}

\maketitle
\section{Introduction}
The modern development of quantum technologies relies on the unique and fundamental properties of physical systems in the quantum regime \cite{Nielsen2010}. These properties, often referred to as quantum resources, have become central to advancing the instrumentation in quantum science and engineering. In recent years, the quantification of quantum resources in physical systems has garnered significant attention since it plays a pivotal role in understanding and exploiting quantum mechanics for practical applications \cite{QRT1,QRT2}. Quantum resource theory provides a robust framework for quantifying these resources, offering elegant and systematic tools to characterize quantum systems. In addition to the established resources, such as entanglement \cite{Bell,EPR,Sch} and coherence \cite{Girolami,Baumgratz2014,Streltsov2017}, the theory has expanded to consider other quantum phenomena, including quantum discord \cite{discord}, non-locality \cite{Luo2011PRL}, and asymmetry \cite{A1,A2,A3,A4,A5,A6,A7}, which are now being recognized as valuable resources for specific quantum tasks. The ability to quantify and manipulate these resources is the key to developing more efficient quantum technologies, such as quantum error correction, quantum communication, and quantum metrology. Moreover, the concept of resource theories has also been extended to broader areas, such as thermodynamics, where quantum coherence and work extraction in quantum systems are studied \cite{Korzekwa2016,Plesnik2024,Skrzypczyk2014}. By understanding the role of these quantum resources, researchers are exploring new avenues for improving the performance and scalability of quantum devices, making quantum technologies more feasible for real-world applications.

 In an attempt to identify promising materials for the implementation of next-generation quantum technologies, researchers have explored a variety of physical qubit systems, including photons \cite{Huang2025}, trapped atoms \cite{Jaksch1998,Bernien2017,Keesling,Long2025}, quantum dots \cite{dots}, nuclear and electron spins, and superconducting circuits \cite{SC1,SC2,SC3,SC4,SC5,SC6}. Recently, fermionic systems have been recognized as viable platforms for physical qubits, with quantum entanglement properties being actively studied in these systems \cite{Ding2025,Jasiukiewicz,WangPRB2025,Kulig2025,Kim2025}. However, each system has its own limitations when it comes to implementing information processing tasks, and no single physical system can be reliably controlled at room temperature. Exploiting and harnessing quantum resources in physical systems at room temperature continues to be a challenging task even today. In this context, the present paper investigates the behavior of quantum resources, such as entanglement, measurement-induced nonlocality (MIN), and coherence, in molecular magnetic (MM) materials. MM materials typically exhibit strong exchange coupling between metal ions and radicals. By exploiting this strong exchange interaction, it is possible to stabilize the intrinsic quantum features of the system, protecting them from decoherence effects. Notably, recent studies have reported the observation of quantum entanglement in various MM materials at relatively higher temperatures \cite{Ghannadan2021,Ghannadan2022,Souza2008,Cruz2016,Reis2012,Muthu1,Cencarikova}. Furthermore, the effects of Dzyaloshinskii–Moriya interaction on quantum resources in different physical systems have been studied and explored in detail \cite{Houca2022,Mogine2023,Adnane2023,Adnane2024,Houca2022A}.

In this article, we investigate the behavior of quantum resources like  entanglement, measurement-induced nonlocality (MIN), and quantum coherence in a mononuclear nickel-based molecular magnet, $(\text{Et}_3\text{NH})[\text{Ni(hfac)}_2L]$. In this complex, HL refers to 2-(2-hydroxy-3-methoxy-5-nitrophenyl)-4,4,5,5-tetramethyl-4,5-dihydro-1H-imidazol-3-oxide-1-oxyl, and hfacH stands for hexafluoroacetylacetone. This system offers a unique platform for the experimental realization of a mixed spin-(1/2, 1) Heisenberg dimer, which is of particular interest in quantum information science. The central $\text{Ni}^{2+}$ ion, with its unpaired electrons, can be modeled as a spin-1 (qutrit) system and effectively manipulated via external magnetic fields or microwave pulses. The nitronyl nitroxide radical, on the other hand, behaves as a spin-1/2 (qubit) system. Together, they form a hybrid spin system, making the compound a compelling candidate for encoding and manipulating quantum information. Crucially, the exchange interaction ($J$) between the $\text{Ni}^{2+}$ ion and the radical is antiferromagnetic, which naturally promotes the formation of entangled states. The  Hamiltonian of the system also includes Zeeman interaction terms, providing tunability and control over the quantum state, an essential feature for implementing quantum logic operations and gate-based computation.

Moreover, the thermal and magnetic resilience of quantum resources in this system indicates that $(\text{Et}_3\text{NH})[\text{Ni(hfac)}_2L]$ offers robustness against environmental decoherence. This makes the compound a promising and practical platform for solid-state quantum technologies, especially in scenarios requiring operation at or near room temperature. By constructing the thermal state of the nickel-based molecular magnet, we study the quantum entanglement, measured using negativity and measurement-induced nonlocality (MIN). We show that entanglement exists in the material at room temperatures. However, while the MIN and coherence survive up to 600 K, negativity ceases to exist above 550 K. Furthermore, we report the influence of the magnetic field on the quantum resources such as entanglement, MIN, and quantum coherence.

The remainder of the paper is organized as follows: In Section \ref{sec2}, we introduce the theoretical model of the hybrid spin system under investigation. In Section \ref{sec3}, we present the quantum correlation quantifiers employed in this study. In Section \ref{sec4}, we analyze the behavior of ground-state and thermal quantum correlations in the mononuclear nickel-based molecular magnet. Finally, in Section \ref{cncl}, we summarize the key results and highlight the significant outcomes of the present investigation.


\section{Model}
\label{sec2}
In this section, we introduce the theoretical model of the molecular complex $(\text{Et}_3\text{NH})[\text{Ni(hfac)}_2L]$. The Hamiltonian of the corresponding theoretical model of the physical system is given by  
\begin{align}
\mathcal{H}=J \Vec{\hat{s}}\cdot\Vec{\hat{S}}-g_{\text{Rad}}\mu_B B \hat{s}^z-g_{\text{Ni}}\mu_B B\hat{S}^z
\label{Hamiltonian}
\end{align}
where $\hat{s}^{\alpha}~(\hat{S}^{\alpha})$  denotes components of the spin-$1/2 ~(1)$ operators ascribed to the nitronyl-nitroxide radical  ($\text{Ni}^{2+}$ magnetic ion) with $\alpha=x,y,z$. The parameters $g_{\text{Rad}}=2.005$ and $g_{\text{Ni}}=2.275$ denote the Lande g-factors corresponding to the nitronyl-nitroxide radical and the $\text{Ni}^{2+}$ ion, respectively, $J$ is the exchange coupling constant between the spin-1/2 and spin-1 magnetic ions.  We wish to point out that  we have adopted natural units with $\hbar = k_B = \mu_B = 1$ throughout the theoretical framework. Energies are measured relative to the exchange scale $J/k_B$, and the magnetic field appears only through the dimensionless ratio $g\mu_B B/J$. 
Although Ni$^{2+}$ ions possess spin–orbit coupling, the ligand field largely quenches the orbital contribution, leading primarily to small g-factor anisotropies rather than significant exchange anisotropy. Consequently, XXZ/XYZ corrections or Ising type models are expected to be much smaller than the dominant isotropic superexchange scale and do not qualitatively affect the thermal quantum correlations studied here. The isotropic Heisenberg model therefore captures the essential physics of the molecular complex. We further neglect the single-ion anisotropy term $D$ acting on the $S=1$ Ni$^{2+}$ ion. For the nickel–radical complex under consideration, the crystal-field anisotropy is expected to be weak compared to the isotropic exchange interaction. 

\begin{figure}[!h]
	\begin{center}
		\includegraphics[width=0.45\textwidth,height=145px]{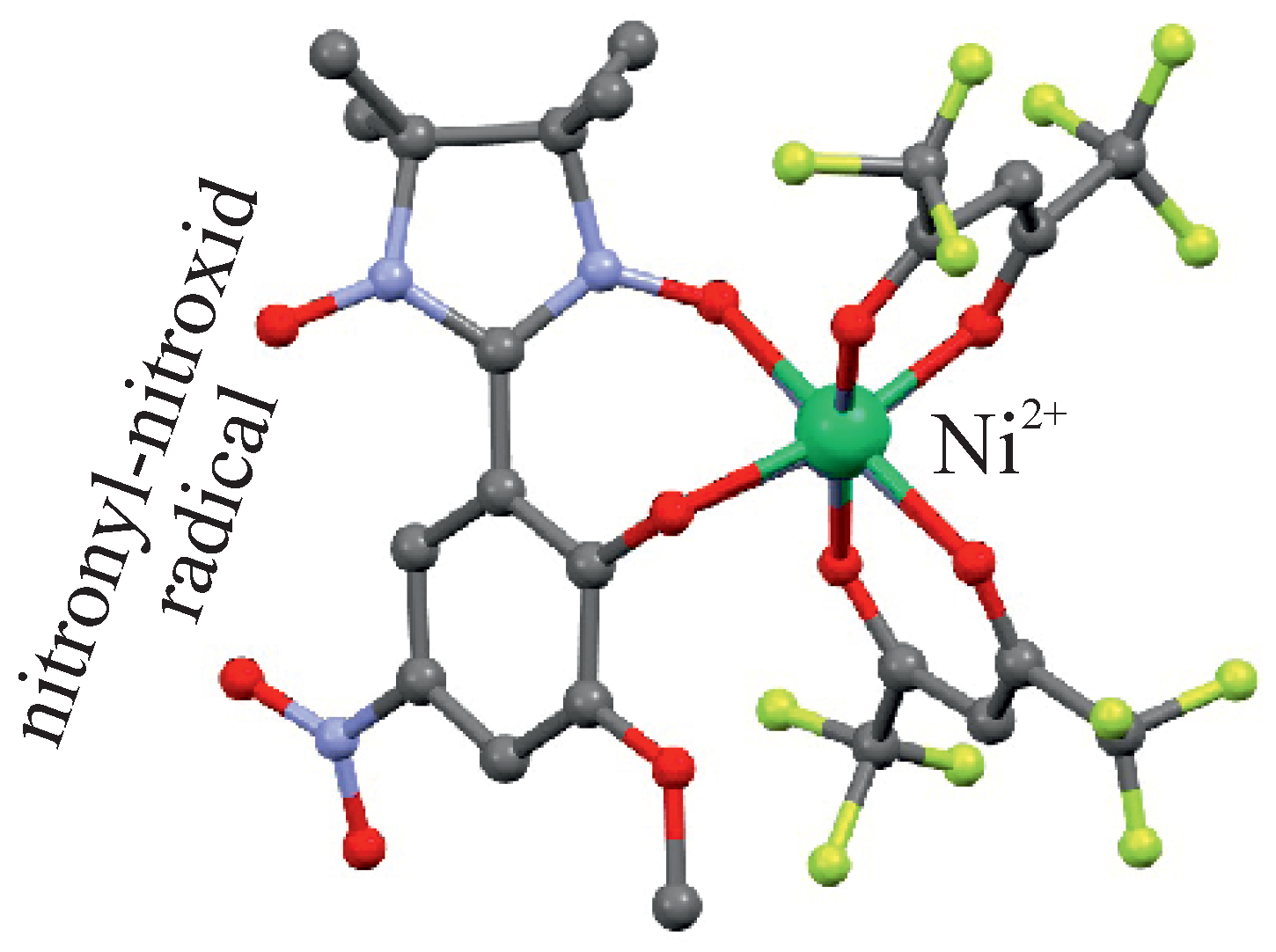}
  		\end{center}
\caption{Schematic representation of the nickel-radical molecular complex $(Et_3NH)[Ni(hfac)_2L]$ \cite{Spinu2021}} 
\label{fig0}
\end{figure}

In the standard computational basis  $\{|\frac{1}{2},1\rangle, |\frac{1}{2},0 \rangle, |\frac{1}{2},-1\rangle, |-\frac{1}{2},1\rangle, |-\frac{1}{2},0\rangle, |-\frac{1}{2},-1\rangle\}$, the Hamiltonian Eq. (\ref{Hamiltonian}) has the following eigenvalues 
\begin{align}  
    \delta_{1,2}=&\frac{1}{2}( J \mp (h_1 + 2h_2))\notag \\
    \delta_{3,4}=&-\frac{1}{4}(J + 2h_2)\mp \frac{1}{4}\sqrt{\eta_-^2+8J^2} \nonumber \\
    \delta_{5,6}=&-\frac{1}{4}(J+2h_2)\mp \frac{1}{4}\sqrt{\eta_+^2+8J^2} \nonumber
\end{align}
where  $\eta_{\mp}=J\mp2(h_1-h_2)$ and the corresponding eigenvectors are 
\begin{align}
   & |\varphi_{1,2}\rangle=|\mp\frac{1}{2},\mp 1 \rangle \nonumber \\ 
    &|\varphi_{3,4}\rangle=\mathcal{\alpha}_{\mp}|\frac{1}{2},0\rangle \mp \mathcal{\alpha}_{\pm} |-\frac{1}{2},1\rangle \nonumber \\
    &|\varphi_{5,6}\rangle=\mathcal{\beta}_{\pm}|\frac{1}{2},-1\rangle \mp \mathcal{\beta}_{\mp} |-\frac{1}{2},0\rangle \nonumber
\end{align}
while the normalization constants are given by
\begin{align}
    \mathcal{\alpha}_{\pm}= \frac{1}{\sqrt{2}}\sqrt{1\pm \frac{\eta_-}{\eta^2_-+8J^2}} ~~~~~~~~~~\beta_{\pm}=\frac{1}{\sqrt{2}}\sqrt{1\pm \frac{\eta_+}{\eta^2_++8J^2}}.
\end{align}
The thermal density matrix of the interacting spin system given by Eq. (\ref{Hamiltonian})  is defined as  
\begin{align}
    \rho(T)=\frac{1}{\mathcal{Z}}\exp{\left(-\beta \mathcal{H}\right)}=\frac{1}{\mathcal{Z}}\sum_{i=1}^6 p_i \vert \varphi _{i}\rangle \langle \varphi _{i}\vert,
\end{align}
where $p_i$ are the eigenvalues of $\rho(T)$, $\beta=1/k_BT$ and $\mathcal{Z}=\text{Tr}\exp{\left(-\beta \mathcal{H}\right)}$ is the partition function and $k_B$ is Boltzmann's constant, which is set to unity for simplicity.  The thermal state of the mixed spin system is calculated as 
\begin{align}
\rho(T) = \frac{1}{\mathcal{Z}}\begin{pmatrix}
 \varrho_{11} & 0 & 0 & 0 & 0 & 0 \\
 0 & \varrho_{22} & 0 & \varrho_{24} & 0 & 0 \\
 0 & 0 & \varrho_{33} & 0 & \varrho_{35} & 0 \\
0 & \varrho_{42} & 0 & \varrho_{44} & 0 & 0 \\
0 & 0 & \varrho_{53} & 0 & \varrho_{55} & 0 \\
0 & 0 & 0 & 0 & 0 & \varrho_{66} \\
\end{pmatrix},
\label{thermal}
\end{align}
with the matrix diagonal elements defined as 
\begin{align}
\varrho_{11}&=\frac{1}{Z}\exp(-\frac{\beta}{2}\chi_-),\;\;\;\;\;\;\;\;\;\;\; \;\;\;\;  \varrho_{66}=\frac{1}{Z}\exp(-\frac{\beta}{2}\chi_+) \nonumber \\
\varrho_{22}&=\frac{1}{Z}\exp[\frac{\beta}{4}(J+2h_2)]\left[ \text{cosh}(\frac{\beta}{4}\eta_-)-\frac{J-2(h_1-h_2)}{\eta_-} \text{sinh}(\frac{\beta}{4}\eta_-)\right] \nonumber \\
\varrho_{33}&=\frac{1}{Z}\exp[\frac{\beta}{4}(J-2h_2)]\left[ \text{cosh}(\frac{\beta}{4}\eta_+)+\frac{J+2(h_1-h_2)}{\eta_+} \text{sinh}(\frac{\beta}{4}\eta_-)\right] \nonumber \\
\varrho_{44}&=\frac{1}{Z}\exp[\frac{\beta}{4}(J+2h_2)]\left[ \text{cosh}(\frac{\beta}{4}\eta_-)+\frac{J-2(h_1-h_2)}{\eta_-} \text{sinh}(\frac{\beta}{4}\eta_-)\right] \nonumber \\
\varrho_{55}&=\frac{1}{Z}\exp[\frac{\beta}{4}(J-2h_2)]\left[ \text{cosh}(\frac{\beta}{4}\eta_+)-\frac{J+2(h_1-h_2)}{\eta_+} \text{sinh}(\frac{\beta}{4}\eta_-)\right] \nonumber  
\end{align}
where $\chi_{\pm}=J+2D\pm(h_1+2h_2)$ and the off-diagonal elements of the thermal state being 
\begin{align}
\varrho_{24}&= \varrho_{42}=-\frac{\sqrt{8}J\exp[\frac{\beta}{4}(J+2h_2)]}{Z\sqrt{\eta_-}}\text{sinh}\left(\frac{\beta}{4}\eta_-\right)\;\;\;\;\text{and}\;\;\;\;\\
\varrho_{35}&= \varrho_{53}=-\frac{\sqrt{8}J\exp[\frac{\beta}{4}(J-2h_2)]}{Z\sqrt{\eta_+}}\text{sinh}(\frac{\beta}{4}\eta_+). \nonumber 
\end{align}
The partition function of the system is given by
\begin{align}
 \mathcal{Z}=2\left[\mathrm{e}^{\frac{-\beta J}{2}}\text{cosh}\left(\frac{\beta(h_{1}+2h_{2})}{2}\right)+\mathrm{e}^{\frac{\beta J}{4}}\left[\mathrm{e}^{\frac{\beta h_{2}}{2}}\text{cosh}\left(\frac{\beta\eta_{-}}{4}\right)+\mathrm{e}^{\frac{-\beta h_{2}}{2}}\text{cosh}\left(\frac{\beta\eta_{+}}{4}\right)\right]\right].
\end{align}
\section{Quantum Correlation Quantifiers}
\label{sec3}
In this section, we review the concepts of quantum correlation measures such as negativity,  measurement-induced nonlocality (MIN) and $l_1-$ norm coherence and  plan to analyze them  in the mononuclear nickel-based magnet $(Et_3NH)[Ni(hfac)_2L]$.  Let $\rho$ be the density matrix defined in the composite Hilbert space $\mathcal{H}_{ab}=\mathcal{H}_{a}\otimes \mathcal{H}_{b}$. The reduced density matrix of the subsystem, $\rho_{a(b)}$, is obtained by performing a partial trace over the other subsystem, that is, $\rho_{a(b)}=\text{Tr}_{b(a)}(\rho)$ and is defined in the Hilbert space $\mathcal{H}_{a(b)}$.
\subsection{Negativity}
Quantum entanglement is an ubiquitous property of composite physical systems in the quantum regime. In recent times, a wide variety of measures have been introduced to characterize entanglement. Among all entanglement measures, negativity is arguably the best-known and most popular tool for quantifying bipartite quantum correlations, particularly in systems of the form $\mathcal{C}^2\otimes \mathcal{C}^3$. Based on the Pere-Herodecki PPT criterion, the measure negativity is introduced by Vidal and Werner, and  is defined as \cite{vidal2002}  
\begin{align}
    \mathcal{N}(\rho)=\sum_{i=1}^6\frac{1}{2}(|\lambda_i|-\lambda_i)
    \label{negativity}
\end{align}
where $\lambda_i$ are the eigenvalues of partially transposed density matrix derived from the density matrix $\rho$ upon partial transposition. The state is entangled if and only if at least one of the eigenvalues is negative; Otherwise, the state is separable \cite{Peres1996,Horodecki1996}. The partially transposed density matrix $\rho^{T_{A}}$ of Eq. (\ref{thermal}) can be written as 
\begin{align}
\rho^{T_{A}} = \frac{1}{\mathcal{Z}}\begin{pmatrix}
 \varrho_{11} & 0 & 0 & 0 & \varrho_{24} & 0 \\
 0 & \varrho_{22} & 0 & 0 & 0 & \varrho_{35} \\
 0 & 0 & \varrho_{33} & 0 & 0 & 0 \\
0 & 0 & 0 & \varrho_{44} & 0 & 0 \\
\varrho_{42} & 0 & 0 & 0 & \varrho_{55} & 0 \\
0 & \varrho_{53} & 0 & 0 & 0 & \varrho_{66} \\
\end{pmatrix},
\label{PPT}
\end{align}
After a straightforward diagonalization of the above partially transposed matrix, the eigenvalues of the matrix are given by

\begin{eqnarray}
    \lambda_1=\varrho_{33}, ~~~~~~~~~~    \lambda_2=\varrho_{44,} ~~~~~~~~~~~~~~~ \\
    \lambda_{3,4}=\frac{\varrho_{22}+\varrho_{66}}{2}\pm \frac{1}{2}\sqrt{(\varrho_{22}-\varrho_{66})^2+4\varrho_{35}\varrho_{53}},  \\
     \lambda_{5,6}=\frac{\varrho_{11}+\varrho_{55}}{2}\pm \frac{1}{2}\sqrt{(\varrho_{11}-\varrho_{55})^2+4\varrho_{24}\varrho_{42}}.
\end{eqnarray}
Substituting the eigenvalues of the partially transposed density matrix in Eq. (\ref{negativity}), we calculate the negativity of the nickel-radical system. 
\subsection{Measurement--Induced Nonlocality}
In the domain of quantum information research community, it is widely believed that entanglement does not fully capture the nonlocal aspects of a bipartite system. To address this, Luo and Fu introduced a new quantum correlation measure from a geometric perspective, aiming to capture the complete nonlocality of a quantum system. This measure is defined as \cite{Luo2011PRL}:
\begin{align}
   \text{ MIN} =~^\text{max}_{\Pi^a} || \rho-\Pi^a(\rho)||^2
\end{align}
where $\Pi^a(\rho)$ is the post-measurement state defined as $\Pi^a(\rho)=\sum_k(\Pi^a_k\otimes\mathds{1}^b) \rho (\Pi^a_k\otimes\mathds{1}^b)$ with $\Pi^a_k$ being the eigenprojectors.  The projectors $\Pi^a_k$ are constructed from the eigenvectors of the subsystem $a$. A key distinction between entanglement and measurement-induced nonlocality (MIN) lies in the effect of local operations. In the case of entanglement (e.g., via partial transpose operations), the subsystems are altered. In contrast, while measuring MIN, the local subsystems remain unchanged. Due to this invariance of the marginal states under local von Neumann measurements, MIN is considered as a more robust and secure resource for quantum information processing tasks compared to entanglement.

The closed formula of MIN for $\mathcal{C}^2\otimes \mathcal{C}^3$ dimensional system is given by
\begin{equation}
\text{MIN} =
\begin{cases}
\text{Tr}(TT^t)- \frac{1}{|\text{x}|^2}\text{x}^tTT^t\text{x}~~~~
 \text{if}~~~~~~~~ \text{x} \neq  0, \\
\text{Tr}(TT^t)-\tau_{\text{min}} ~~ ~~~~~~~~\text{otherwise}.
\end{cases} 
\label{HSMIN}
\end{equation}
where $\tau_{\text{min}}$ is the least eigenvalue of matrix $TT^t$  and the matrix elements of the correlation matrix $T$ are defined as $t_{ij}=\text{Tr}[\rho(\sigma_i\otimes \Lambda_j)]$. Here, $\sigma_i$ and $\Lambda_j$ are the spin-1/2 and spin-1 operators. 
\subsection{$l_1-$norm of coherence}
$l_1-$norm of coherence, a widely used quantifier of quantum coherence in the resource theory of coherence, introduced by Baumgratz et al. in 2014, measures the amount of superposition (or off-diagonal elements) in a given quantum state with respect to a fixed reference basis. For the thermal state  $\rho(T)$  represented in a fixed orthonormal basis, the $l_1-$norm of coherence is defined as \cite{Baumgratz2014}
\begin{align}
    C_{l_1}(\rho(T))=\sum_{i\neq j} |\varrho_{ij}|
\end{align}
where $\varrho_{ij}$ are the off-diagonal elements of the density matrix $\rho(T)$. 
\section{Results and Discussion}
\label{sec4}
In the following, we investigate the quantum resource content of the nickel–radical molecular complex $(\text{Et}_3\text{NH})[\text{Ni(hfac)}_2L]$ in its ground state as well as in the corresponding thermal quantum state.
\subsection{Ground-state entanglement}
It is worth noting that all pure entangled states are nonlocal, implying that entanglement is sufficient to characterize the nonlocal aspects of the pure state. Before presenting a detailed analysis of thermal quantum correlations in the Nickel-Radical Molecular Complex $(\text{Et}_3\text{NH})[\text{Ni(hfac)}_2L]$, we first examine the quantum resource content in the ground state of the system under our consideration. In the absence of an external magnetic field, the ground state energy of the nickel-radical system is $E_3=E_5=-J$, indicating a two-fold degeneracy. The corresponding ground-state density matrix is a coherent superposition state and is given by
\begin{align}
\rho(0)=\frac{1}{2}[(|\varphi_3\rangle +|\varphi_5\rangle)(\langle \varphi_3|+\langle \varphi_5|)]
\end{align}
which represents a coherent superposition within the degenerate ground-state subspace. Notably, the system exhibits non-zero entanglement at zero temperature, with the calculated negativity approximately being 0.33.  Due to the intervention of the magnetic field, the two-fold degeneracy of the ferrimagnetic doublet is removed through Zeeman splitting of the energy levels. Then, the ground state of the system is given by 
\begin{align}
    \rho(0)=|\varphi_{3}\rangle \langle \varphi_{3}|
\end{align}
and the ground state entanglement is calcultated as
\begin{align}
    \mathcal{N}(\rho(0))=\frac{\sqrt{2}J}{[J+B(g_{\text{Ni}}-g_{\text{Rad}})]^2+8J^2}.
\end{align}
It is clearly evident that the ground state entanglement is a function of the interaction strength, the difference of g-factor and magnetic field. For the system under consideration, the parameters are $J/k_B=505 K$, $g_{\text{Ni}}=2.275$, and $g_{\text{Rad}}=2.005$. Consequently, the entanglement of the system is inversely proportional to the magnetic field.

\begin{figure}[!h]
	\begin{center}
		\includegraphics[width=0.3\textwidth, height=120px]{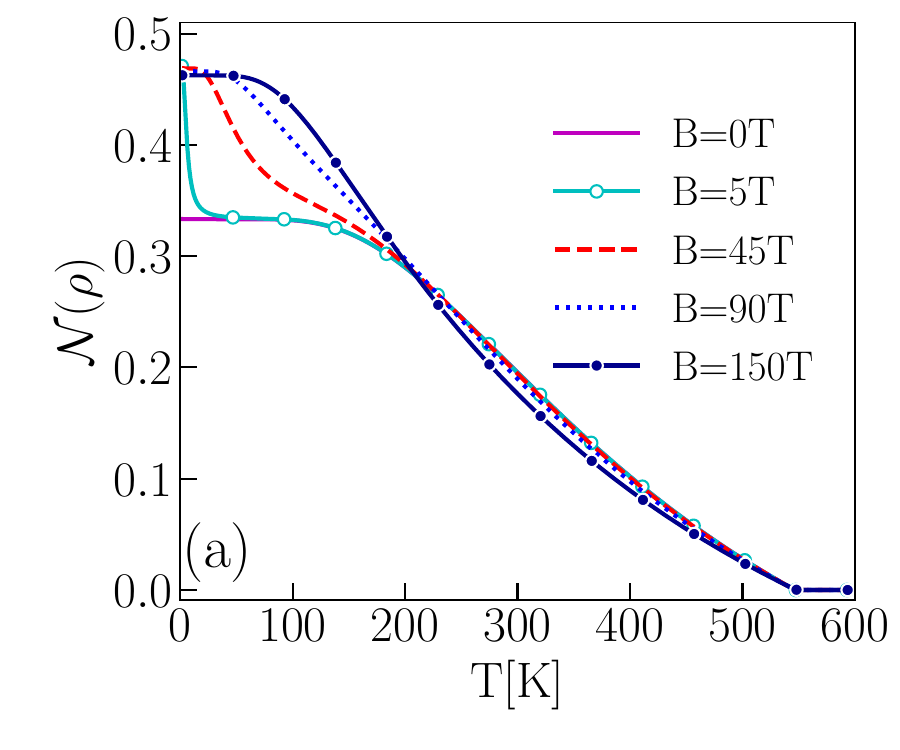}
		\includegraphics[width=0.3\textwidth, height=120px]{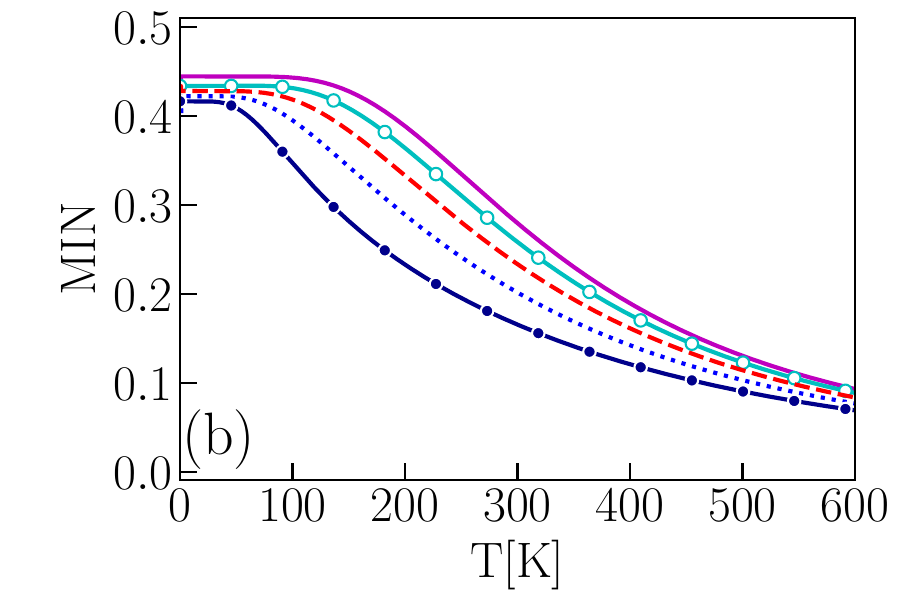}
        \includegraphics[width=0.3\textwidth, height=120px]{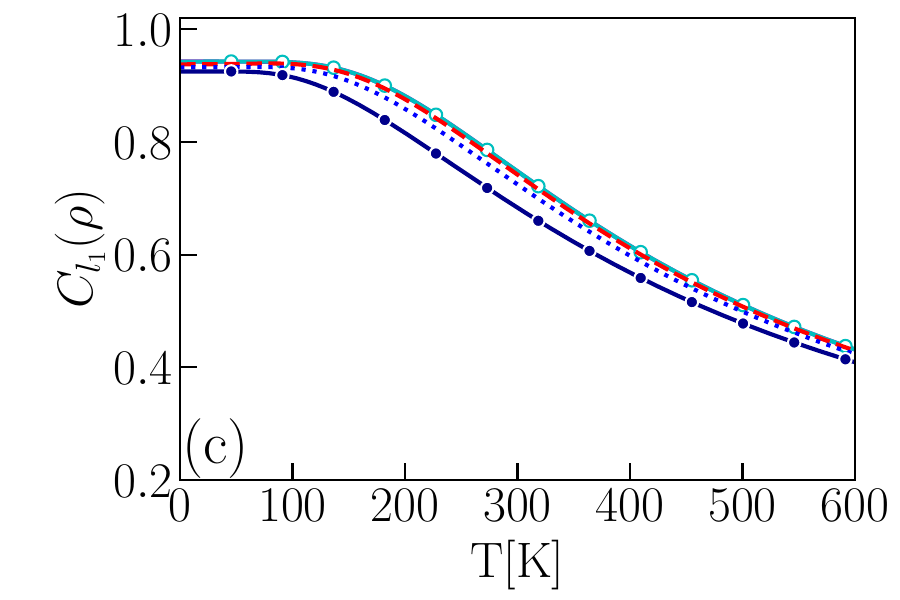}
  		\end{center}
\caption{ Variations of quantum resources such as (a) negativity (b) MIN and (c) $l_1-$norm of coherence as a function of temperature at a few selected values of the external magnetic field for the nickel-radical molecular compound described by the Hamiltonian (1) with the parameters set $J/k_B = 505 K$, $g_{Rad} = 2.005$, and $g_{Ni} = 2.275$.} 
\label{fig1}
\end{figure}

\subsection{Thermal quantum resources}
Next, we investigate the behavior of quantum resources such as negativity (a measure of entanglement), measurement-induced nonlocality (a measure of quantum correlations beyond entanglement), and coherence quantified by $l_1-$norm in the molecular complex $(\text{Et}_3\text{NH})[\text{Ni(hfac)}_2L]$ at nonzero temperatures.  In this context, we model the theoretical Hamiltonian of the nickel-radical molecular compound as a mixed spin-$(1/2,1)$ dimer, using the set of specific parameters: $J/k_B=505 K$, $g_{\text{Ni}}=2.275$, and $g_{\text{Rad}}=2.005$. 

To illustrate the behavior of quantum correlations and coherence in the system under consideration, we have plotted the negativity, measurement-induced nonlocality (MIN), and coherence in Fig. (\ref{fig1}) as functions of temperature for selected values of the external magnetic field. As shown in Fig.(\ref{fig1}a), negativity exhibits a non-monotonic dependence on temperature across all magnetic field strengths. At zero magnetic field (B = 0 T), the system  possesses a finite amount of entanglement at low temperatures, which gradually diminishes with increasing temperature and vanishes beyond approximately at $T \approx 550 K$. This decay is attributed to thermal mixing, which dilutes the coherent quantum superpositions responsible for entanglement. This result is identical to the previously reported value for the same model \cite{Inorganics2024}. In the low-temperature limit, the behavior of negativity demonstrates a sensitive dependence on weak magnetic fields, indicating that small field strengths can still sustain entanglement within a narrow thermal range. When $T \approx 250 K$,  the entanglement curves for all magnetic fields merge into a single line.  Furthermore, for $T \approx 250 K$, the entanglement is suppressed solely due to thermal fluctuations, with no influence from the magnetic field. As the magnetic field strength increases, the overall magnitude of negativity decreases, and the temperature range supporting non-zero entanglement in the parameter space becomes narrower. For example, at a high magnetic field $(B=150T)$, entanglement is almost completely suppressed, even at low temperatures. This suggests that a strong external magnetic field destroys quantum entanglement by energetically splitting the spin states and disrupting the entangled configurations.

On the other hand, Figure (\ref{fig1}b) shows the variation of the other companion measure, namely, measurement-induced nonlocality (MIN), as a function of temperature for the same set of magnetic fields. Unlike negativity, MIN exhibits a monotonic decrease with increasing temperature at each given magnetic field. At $B=0T$, MIN begins at a relatively high value at low temperatures and gradually decreases with temperature, demonstrating a more robust behavior than negativity. While the entanglement quantified by the negativity vanishes around $T\approx 550K$, the quantum correlations measured by MIN remain non-zero even beyond $T>550 K$.  This observation confirms that quantum correlations beyond entanglement can persist even when entanglement is no longer present. Notably, even under strong magnetic fields (e.g., B = 90 T and 150 T), MIN retains finite values over a broad temperature range, highlighting its resilience to both thermal fluctuations and external magnetic field effects.  In Figure~(\ref{fig1}c), we illustrate the behavior of thermal coherence $C_{l_1}(\rho)$ as a function of temperature. For all fixed values of the magnetic field, we observe a  trend similar to what is being observed in the measurement-induced nonlocality (MIN),  shown in Figure~(\ref{fig1}b).  Compared to negativity, both MIN and $l_1-$norm of coherence exhibit a more gradual decline and a broader thermal survival range, highlighting their roles as stronger and more reliable indicators of quantum correlations in thermally activated systems.

\begin{figure}[!h]
	\begin{center}
		\includegraphics[width=0.3\textwidth, height=125px]{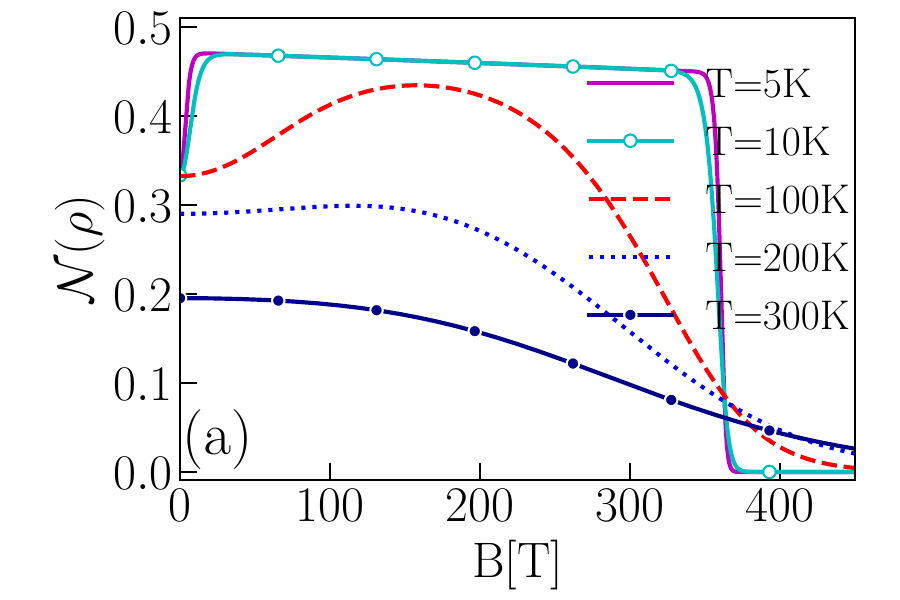}
		\includegraphics[width=0.3\textwidth, height=125px]{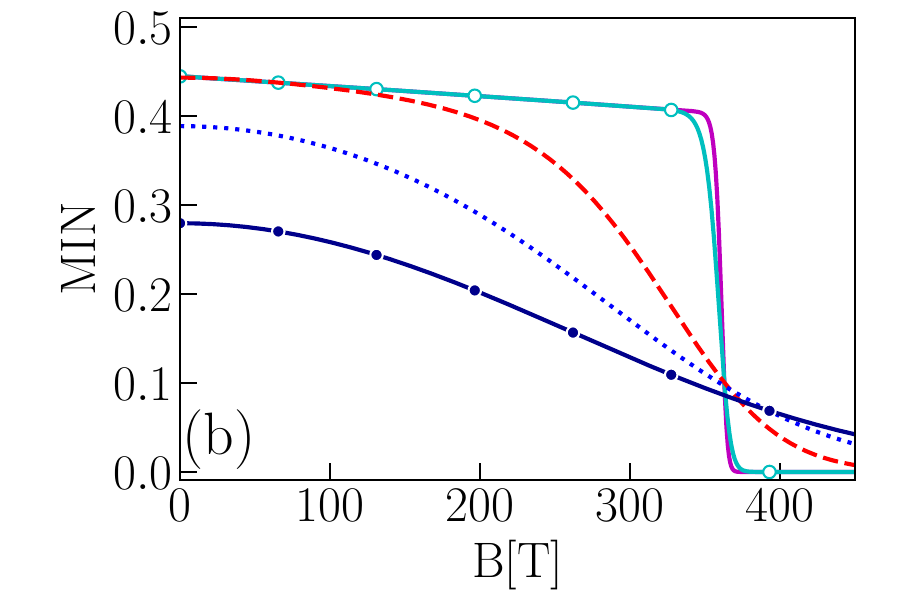}
        \includegraphics[width=0.3\textwidth, height=125px]{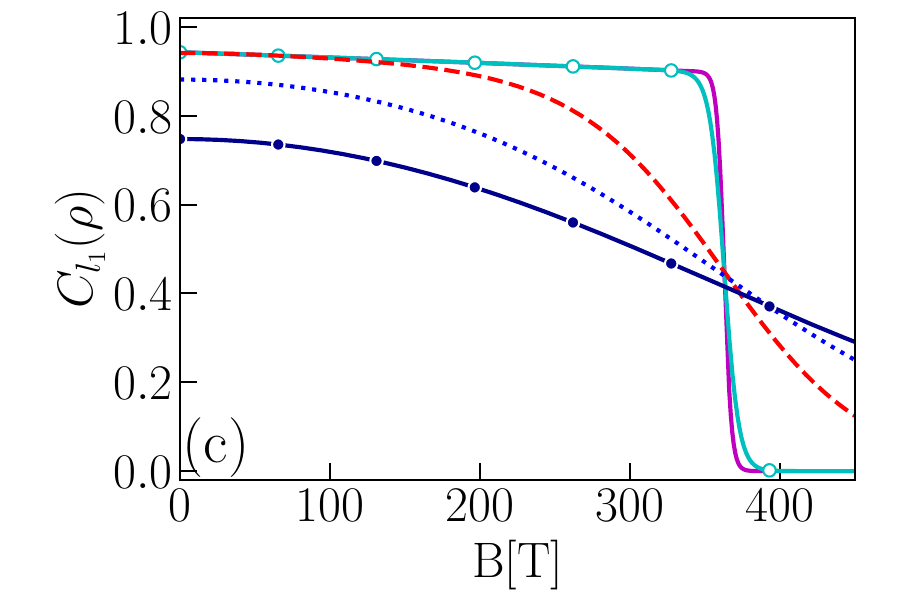}
  		\end{center}
\caption{Variations of quantum resources such as (a) negativity (b) MIN and (c) $l_1-$norm of coherence as a function of magnetic field at a few selected values of the temperature for the nickel-radical molecular compound given by the Hamiltonian (1) with the parameters set $J/k_B = 505 K$, $g_{Rad} = 2.005$, and $g_{Ni} = 2.275$.}
\label{fig2}
\end{figure}

Figure \ref{fig2} illustrates the behavior of negativity, measurement-induced nonlocality (MIN) and quantum coherence as a function of the external magnetic field for various temperatures in the molecular complex $(\text{Et}_3\text{NH})[\text{Ni(hfac)}_2L]$.  The magnetic-field dependence enters through the dimensionless ratio $g \mu_B B/J$. While the experimentally accessible regime corresponds to magnetic fields $B\leq 10~\text{to}~45 T$, we extend the analysis to higher fields merely to determine the theoretical phase boundary of the system. Figure \ref{fig2}(a) displays how negativity varies with the external magnetic field at different temperatures.  At low temperatures (e.g., $T=5K~\&~ 10K$), the negativity begins to increase initially before reaching a plateau like structure where it remains almost a constant and this scenario continues till the magnetic field reaches $\approx 370T$. At $\approx370T$, the negativity suddenly drops to zero. This reflects the field-induced suppression of entanglement due to the energetic separation of spin states. At a critical field $B_c(T)$, the negativity abruptly vanishes. From the above, it is also obvious that the sensitivity to the magnetic field is more pronounced at higher temperatures ($T\geq 100K$).  As the temperature rises, the maximum value of negativity decreases, and its sensitivity to the magnetic field becomes more pronounced. This indicates that thermal fluctuations and magnetic field effects work together to weaken entanglement. For higher magnetic field strengths, the entanglement increases gradually at first, then slowly decreases, which contrasts with the behavior observed at lower temperatures. At higher temperatures, negativity remains almost zero across the higher magnetic field range, suggesting that entanglement is completely suppressed by thermal noise, regardless of the external field strength.  The sharp jumps observed in Fig.~3(a) at low temperatures ($T = 5$~K and $10$~K) arise from magnetic-field–induced level crossings in the energy spectrum of the mixed spin-(1/2,1) Heisenberg dimer. In the low-temperature regime, the thermal state is dominated by the ground state, and a critical magnetic field $B_c$ induces an abrupt change in the ground-state configuration due to Zeeman splitting. This sudden rearrangement of the energy levels leads to discontinuities in the eigenvalues of the partially transposed density matrix, manifesting as jumps in the negativity. As the temperature increases, thermal population of excited states becomes significant, which evens out the level-crossing effects and results in a continuous dependence of the quantum correlations on the magnetic field.

Figure \ref{fig2}(b) shows the variation of MIN with magnetic field for the same set of temperatures. At zero magnetic field, a non-zero correlation exists between $Ni^{2+}$ and the nitronyl-nitroxide, similar to entanglement. This behavior is observed across all the selected temperatures. At  $T= 5K ~\text{and} ~10K$, MIN decreases gradually up to a critical magnetic field, beyond which it drops abruptly to zero. At higher temperatures, unlike negativity, MIN decreases monotonically yet retains a non-zero value even at strong magnetic fields and elevated temperatures. These trends further support the interpretation that quantum correlations beyond entanglement, as quantified by MIN, persist under conditions where entanglement vanishes, making MIN a more robust and reliable quantum resource in thermally active scenarios. The $l_1-$norm of coherence (\ref{fig2}c) exhibits a trend similar to MIN, with a smooth and continuous decay as the magnetic field increases. It shows strong resilience to both temperature and magnetic field variations, capturing persistent off-diagonal coherence elements in the density matrix even when entanglement vanishes. While negativity captures the entanglement dynamics and is highly sensitive to external magnetic field, both MIN and the $l_1-$norm of coherence demonstrate greater robustness. They exhibit a broader survival range under thermal and magnetic disturbances, making them stronger and more reliable resources under the strong magnetic fields. This clearly supports the view that quantumness measures (such as MIN and coherence) beyond entanglement are valuable resources for practical quantum technologies operating at or near room temperatures.

\begin{figure}[!h]
	\begin{center}
		\includegraphics[width=0.3\textwidth, height=145px]{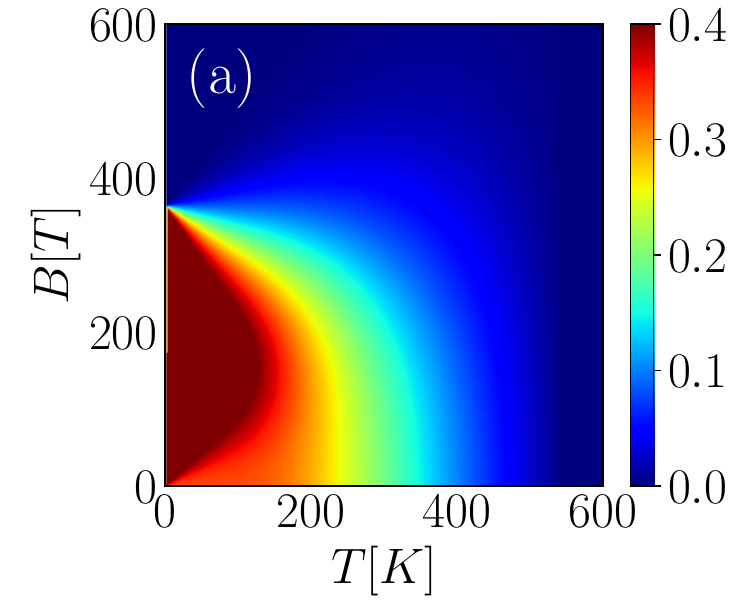}
		\includegraphics[width=0.3\textwidth, height=145px]{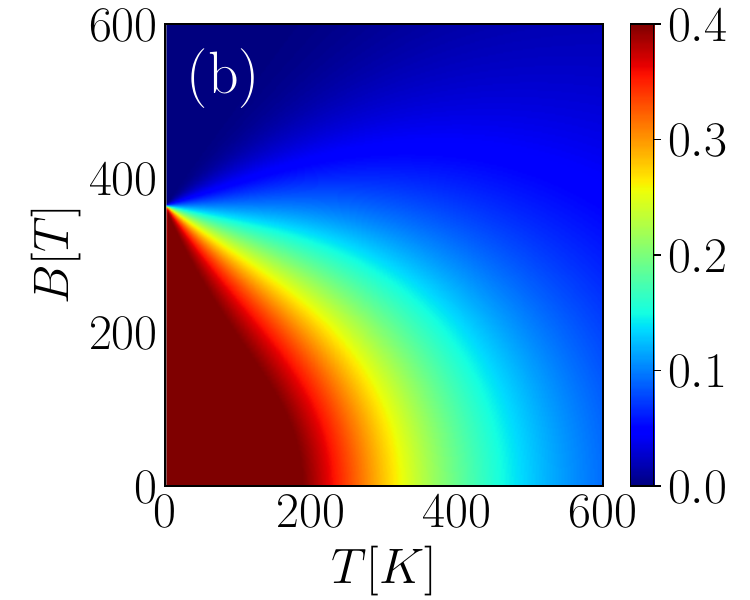}
        \includegraphics[width=0.3\textwidth, height=145px]{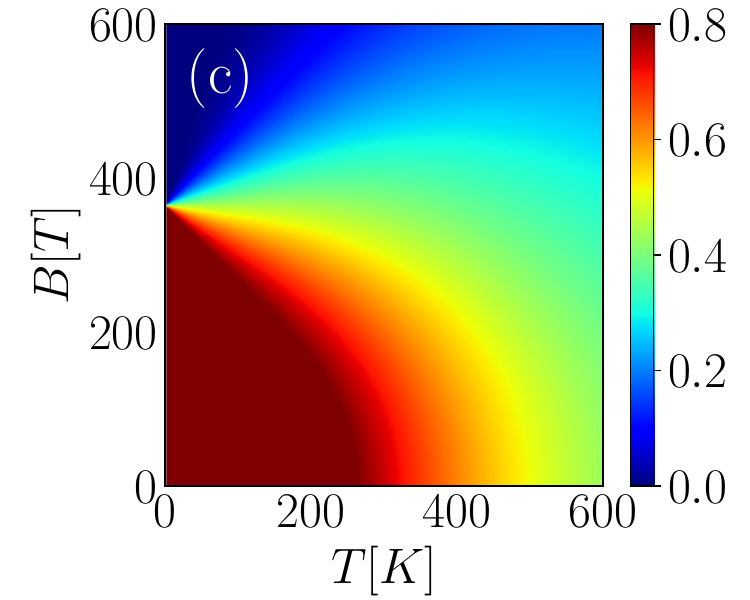}
  		\end{center}
\caption{Density plots of the (a) negativity, (b) MIN, and (c) $l_1-$norm of cohrence in the temperature versus magnetic field plane for the mixed spin-( 1/2 ,1) Heisenberg dimer given by the Hamiltonian (1) with the parameter set $J/k B = 505 K$, $g_{Rad} = 2.005$, and $g_{Ni} = 2.275$ adjusted to a theoretical modeling of the molecular complex $(Et_3 NH)[Ni(hfac)_2 L]$. (a) shows a detail of the density plot for the negativity), while (b) shows a complete density plot up to magnetic fields above which the negativity vanish} 
\label{fig3}
\end{figure}
Figure 4 presents density plots of (a) negativity, (b) measurement-induced nonlocality (MIN), and (c) $l_1-$norm of coherence as functions of temperature $T$ and external magnetic field $B$ for the nickel-based molecular magnet $(Et_3NH)[Ni(hfac)_2L]$.  Figure \ref{fig3} illustrates how  the entanglement strength (Negativity), MIN and coherence vary over a temperature range from $0 ~\text{to}~ 600 K$ and magnetic fields ranging from $0~ \text{to}~ 450 T$. Figure \ref{fig3}a shows the entanglement landscape across the $T-B$ plane. Negativity exhibits high values (dark regions) at low temperatures and low magnetic fields, indicating a strong entangled ground state. As temperature or magnetic field increases, entanglement rapidly decays and completely vanishes beyond critical thresholds $T=550K~~ \text{and}~~ B=370T$.  This sharp decline highlights the fragile nature of entanglement against thermal agitation and field-induced Zeeman splitting. Figure \ref{fig3}b displays the corresponding density map of MIN. Unlike negativity, MIN retains significant values over a much broader range of temperatures and magnetic fields, persisting even where entanglement is zero. This suggests that MIN captures nonclassical correlations beyond entanglement, which survive under extreme conditions. Figure 4c illustrates the behavior of thermal coherence quantified by the $l_1-$norm. Similar to MIN, coherence shows strong values in the low$-T$, low$-B$ regime and decays continuously with increasing $T$ and $B$, but remains non-zero over a wide range. The smooth gradient indicates that coherence is also less susceptible to decoherence compared to entanglement.

\section{Conclusions}
\label{cncl}
To summarize, we have investigated the behaviors of quantum resources in the thermal states of the mononuclear nickel-based magnet $(\text{Et}_3\text{NH})[\text{Ni(hfac)}_2L]$ quantified by negativity, measurement-induced nonlocality (MIN), and $l_1-$norm of coherence. Our analysis, grounded in experimentally obtained spin-Hamiltonian parameters, demonstrates that entanglement is highly sensitive to thermal and magnetic fluctuations and disappears beyond critical thresholds. This indicates that while bipartite entanglement quantified by the negativity vanishes, quantum coherence  which is a more general quantum resource remains stable in a wider parameter regime. In contrast, both MIN and coherence show extended survival across a broad range of temperatures and magnetic fields, making them more resilient indicators of quantum correlations in realistic, noisy environments.  Notably, MIN remains non-zero even at high temperature and strong magnetic fields, reinforcing its potential as a robust quantum resource.

The comparative study underscores the importance of quantum correlations beyond entanglement and their potential applications in thermally activated solid-state systems. These results open up new directions for utilizing molecular magnets in quantum technologies, particularly across situations where maintaining entanglement is challenging. Future investigations could focus on extending these results to multi-qubit networks and exploring operational tasks that harness the thermal resilience of such quantum resources.

\acknowledgments
RR wishes to acknowledge financial assistance received from DST-CURIE (DST-CURIE-PG/2022/54) and \\ANRF  (DST-CRG/2023/008153).

\onecolumngrid
\appendix

\end{document}